\documentclass{article}

\usepackage{arxiv}

\usepackage[utf8]{inputenc} 
\usepackage[T1]{fontenc}    
\usepackage{hyperref}       
\usepackage{url}            
\usepackage{booktabs}       
\usepackage{amsfonts}       
\usepackage{nicefrac}       
\usepackage{microtype}      
\usepackage{lipsum}
\usepackage{graphicx}
\usepackage{amsmath}
\graphicspath{ {./images/} }

\title{Controlled-Variable Selection based on Chaos Theory for the Tennessee Eastman Plant}

\author{
 Sergio F. Yapur \\
  Facultad de Ingenier\'ia Qu\'imica \\
  Universidad Nacional del Litoral\\
  Santiago del Estero 2829 (3000) Santa Fe \\
  \texttt{syapur@fiq.unl.edu.ar} \\
}

\begin{document}
\maketitle
\begin{abstract}
This work explores a link between chaotic signals and the selection of controlled variables for plantwide control system design. Some results are shown for the Tennessee Eastman plant, which is well-known for being a challenging process in the field of plant-wide control. This article provides a systematic, data-driven method to select which variables should be controlled. However, since plantwide control problems are inherently complex, this work does not intend to provide a definite solution, but a complementary analysis to take into account towards the final control system design. The discussion highlights the potential hidden in the chaos theory to reduce the complexity of the resulting control system. 
\end{abstract}

\keywords{Plantwide Process Control \and Chaos Theory \and Tennessee Eastman \and Data-Driven \and Neural Networks }

\section{Introduction}
Plantwide Process Control (PWC) system design is an ongoing research field. In only a few words, this field treats the whole process at once instead of as a collection of individual unit operations. The advantage of this approach is that interactions between process units are considered since the very first analysis, enabling a better decision-making through the control design. 

A major issue is that real plants are often nonlinear and can have a massive amount of inputs and outputs, i.e., an oil refinery, with thousands of variables involved in highly nonlinear phenomena. Such a high complexity can dim the benefits of linear control methods. First off, obtaining a linear model around an operative point can be computationally costly. Even after getting such a model, there is no guarantee that the model will remain representative of the real operation of the industrial site, since in practice is rather impossible to keep all variables at a specific point. Additionally, some linear representations are strongly ill-conditioned in the frequency range of interest. As a consequence, most of the analysis and design methods that are usually accepted in the literature are no longer suitable nor trustworthy. These characteristics suggest that nonlinear concepts may be useful to the control design process, at least to some extent.

The methods from PWC are usually presented as a series of steps, from the degree-of-freedom assessment of the plant up to the controller tuning \cite{Luyben1998, Skogestad_PWC_1}. Some of these steps aim to gain insight into the plant, such that the final control structure is at least reasonable for the entire plant. In this regard, a key decision for the control structure is the selection of the controlled variables (CV). These are the variables  that are fed back into the control system, in order to keep them around desired setpoint values. This work proposes a new kind of application of a chaos characterization to suggest which variables might be critical to control.

\section{Chaos as a Control-System Feature}
\label{sec:headings}

In the context of nonlinear dynamics, chaos is a phenomenon by which the system output is extremely sensitive to initial conditions \cite{Slotine}. This leads to unpredictability of the system output, an essential feature by which chaos is commonly recognized. 

Chaos must be distinguished from random behavior, since chaotic systems are deterministic, not stochastic. The unpredictability of chaotic phenomena is deeply rooted in the relations of the system variables. These interactions amplify any infinitesimal difference between two different initial conditions to produce two completely different system outputs in the long run.

Chaotic systems are a subset of nonlinear systems, meaning that only nonlinear systems are able to present chaotic behavior. What is more, this feature is usually found in strongly nonlinear systems. Some examples are atmospheric dynamics, turbulent flow, buckled elastic structures, mechanical systems with play or backlash, and feedback control devices \cite{Slotine}.

As one might expect, it is always better to avoid chaotic modes in controls systems, especially in the context of PWC design. However, this is nearly an impossible task for high dimensional systems, as there is no way to recognize beforehand which variables are chaotic. A suitable workaround is to change the design procedure into one that tackles the most complex behavior of the plant in the first place. 

\subsection{On Stability and Chaos} 

Another reason to consider chaos theory in control systems is the close relation between chaos and stability, a fundamental property of control systems theory.

Consider the Lyapunov stability of dynamical systems, which establish that an equilibrium point $x^*$ is stable if solutions that initialize near from $x^*$ remain in a neighborhood of $x^*$ at all times. This concept is commonly used in optimal control and Model Predictive Control, to name a few \cite{Tewari}. This concept of stability will be used throughout the remainder of this work.

In chaos theory, the Lyapunov exponent $\lambda$ measures the sensitivity of a given dynamical system to small changes in initial conditions. A natural link between the Lyapunov exponent and stability is given by the sign of the exponent $\lambda$. Whenever $\lambda > 0$ the trajectories of solutions starting nearby will eventually diverge, revealing instability in the Lyapunov sense. More precisely, let $x_0$ be an initial condition, $\Delta x_0$ a arbitrarily small perturbation in $x_0$, and $\Delta x(x_0,t)$ the euclidean distance at time $t$ between the trajectories starting from $x_0$ and from $x_0 + \Delta x_0$. With these definitions, the Lyapunov exponent is given by:

\begin{equation}
    \label{eq:LyapunovExp}
    \lambda = \lim_{t \rightarrow \infty} \ln \frac{|\Delta x(x_0,t)|}{|\Delta x_0|}
\end{equation}

\begin{figure}[ht]
  \centering
  \makebox[\textwidth][c]{\includegraphics[width=0.8\textwidth]{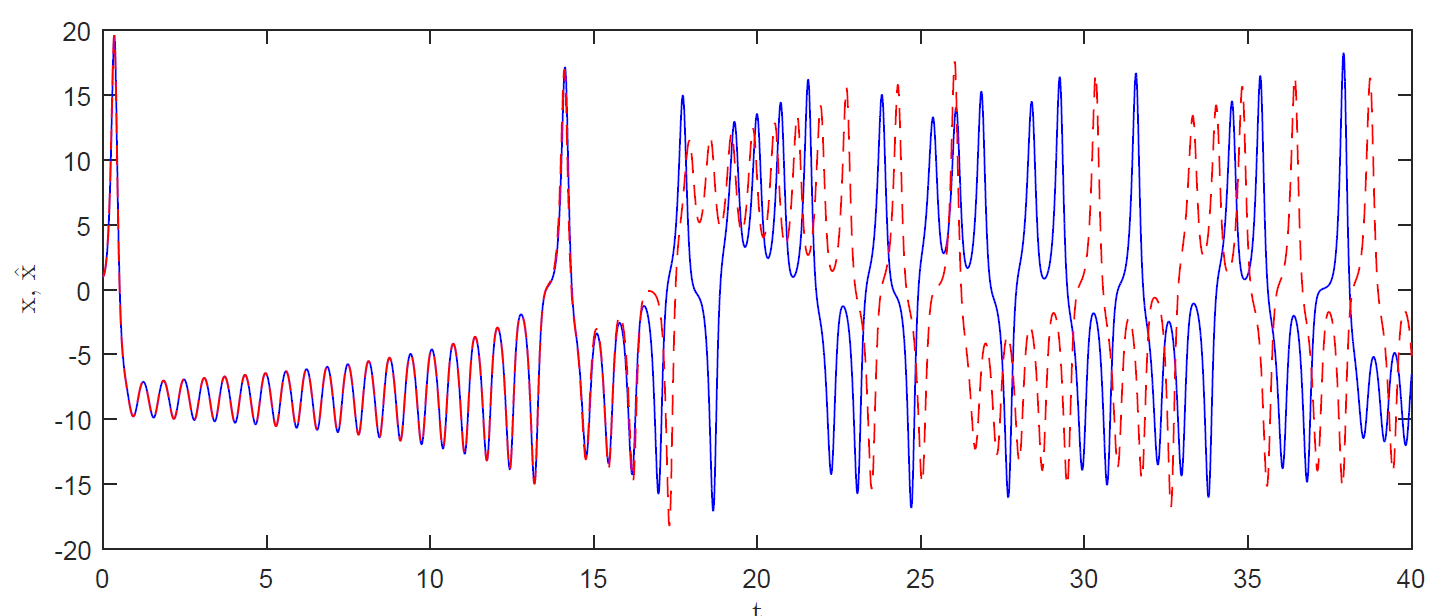}}
  \caption{Lorenz's Model. Dynamic evolution of initially close trajectories. }
  \label{fig:Lorenz}
\end{figure}

Thus, at some time $t$ the separation $d = |\Delta x(x_0,t)|$ of the trajectories is $d \approx d_0 e^{\lambda(t-t_0)}$, where $d_0 = |\Delta x_0|$. Hence chaotic systems have $\lambda >0$, while stable non-chaotic systems have $\lambda<0$. The Fig. \ref{fig:Lorenz} shows a typical dissociation of dynamic trajectories for the Lorenz's model, one of the first studied in the literature.

Typically, a chaotic system will have at least an equilibrium point that is unstable. Therefore, chaoticity implies instability, at least in a local manner. Whenever the plant operative conditions are nearby an unstable equilibrium point, the trajectory complexity will increase arbitrarily. This fact constitutes the cornerstone that supports this exploration.

\subsubsection{Estimation of Chaotic Signals} 

For small and noisy data sets, the verification of chaos behavior is not trivial, let alone the computation of Lyapunov exponents. In either assessment, random variations interfere with the estimation process. Consequently, tests for chaos are scarce in the literature. Recently, a data-driven approach proposed by BenSa\"ida \cite{BenSaida} admits noisy signals. This is possible due to the remarkable advantage of distinguishing between chaos and randomness. The algorithm to detect chaos in signals has two stages. Primarily, the code computes the largest Lyapunov exponent $\lambda$ by using a Jacobian optimization method. Then, a hypothesis test is carried out to assess the presence of chaos in a probabilistic sense. This one-tailed test has a null hypothesis given by $\lambda \ge 0$, which indicates the presence of chaos. The significance level $\alpha$ can be set up to any desired level. The main outputs of the code are the exponent $\lambda$ and the $p-$value. A small value of $p$ casts doubt on the validity of the null hypothesis of chaos.

Internally, the algorithm follows a theoretical formulation given by Eckmann and Ruelle \cite{Eckmann}. Departing from a scalar time series ${x_t}_{t=1}^{T}$, a noisy chaotic system can be written as:

\begin{equation}
    \label{eq:basicSystem}
    x_t = f(x_{t-L},x_{t-2L},...,x_{t-mL}) + \epsilon_t
\end{equation}

where $\epsilon_t$ represents the added noise, and $L$ is a time delay to allow the possibility of skipping samples in the algorithm. The parameter $m$ is the embedding dimension or the length of past dependence. Moreover, the system dynamics is unknown, so it is the following mapping:

\begin{equation}
    \label{eq:mapping}
    F: \left[ \begin{array}{c}
         x_{t-L}  \\
         x_{t-2L} \\
         \vdots \\
         x_{t-m L}
    \end{array} \right] \rightarrow
    \left[ \begin{array}{c}
         x_{t}  \\
         x_{t-L} \\
         \vdots \\
         x_{t-(m-1) L}
    \end{array} \right]
\end{equation}

The Jacobian of the former mapping offers a consistent estimation of Lyapunov exponents, even in the presence of noise. However, an approximation of the mapping $F$ is required in the first place. Neural networks are suitable for this purpose, since they can approximate any smooth, nonlinear function to any desired accuracy, whenever the complexity of the network is appropriate \cite{Ellacott}.

\begin{figure}[ht]
  \centering
  \makebox[\textwidth][c]{\includegraphics[width=1.0\textwidth]{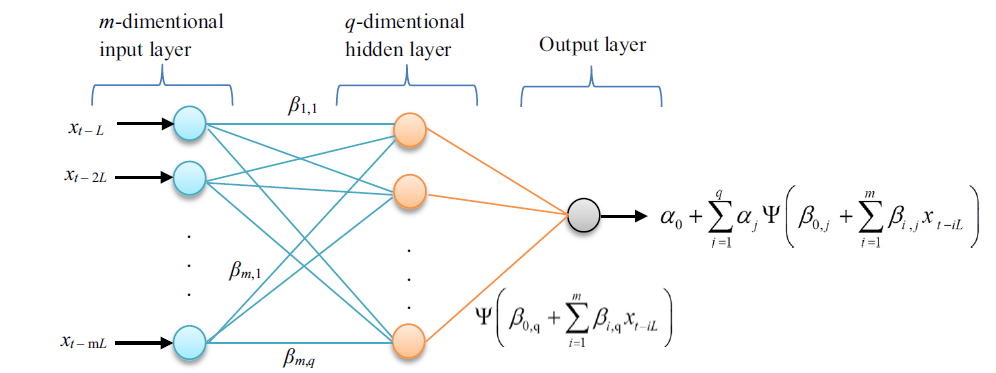}}
  \caption{A neural network with one hidden layer.}
  \label{fig:NN}
\end{figure}

A relatively simple neural network architecture is enough for this application, with an m-dimensional input layer, a q-dimensional hidden layer, and an output layer, which is shown in Fig. \ref{fig:NN}. The activation function $\Psi$ governs the propagation of the signal to the next layer. The hyperbolic tangent can handle numerical problems better than other commonly used activation functions, being so a natural choice for this work hereafter.

The neural network ''learns'' from the inputs and outputs of the chaotic map $F$ by adjusting its internal coefficients $\alpha_j, \beta_{ij}$. Eventually, the network becomes a reasonable approximation of $F$. For this to happen, the noise should be removed from the data. Consequently, an appropriate estimation algorithm is the nonlinear least squares, which minimizes the noise of the system given by $S(\theta) = \sum_{t=m L+1}^{T}\epsilon^2$, where $\theta$ represents the coefficients vector that define the neural network.

The triplet $(L,m,q)$ accounts for the complexity of the neural network, thus defining the mapping approximation. The time delay $L$ should not be too large, since it would cause loss of information from the signal. Conversely, a small value of $L$ interferes with the distinction between noise and signal. Additionally, the embedding dimension $m$ should be large enough to capture the chaotic dynamics. A method to derive this parameter is by setting to the largest eigenvalue of $X^T X$, where $X$ is the lagged matrix associated with the time series $\{x_t\}$. Any value of the triplet can lead to unnecessary overhead in the numerical computation if chosen too large.

From the mapping $f$ it is possible to estimate the Lyapunov exponent through the expression

\begin{equation}
    \hat{\lambda} = \lim_{M\to \infty} \frac{\ln(v_1)}{2M}
\end{equation}

where $M$ is an integer associated with the evaluation points, $M\approx T^{2/3}$, and $v_1$ stands for the largest eigenvalue of the matrix $T_M' T_M$, with $T_M = \prod_{t=1}^{M-1} J_{M-t}$ and 
\begin{equation}
    \label{eq:Jacobiano}
    J_t = 
    \begin{pmatrix}
    \frac{\partial f}{\partial x_{t-L}} &...& \frac{\partial f}{\partial x_{t-m L}} \\
    1 & ...& 0 \\
    \vdots & \ddots & \vdots \\
    0 & ... & 1\\
    \end{pmatrix}
\end{equation}

Additionally, a practical procedure to confirm chaos with high probability is to retrieve positive $\lambda$ for different choices of the triplet $(L,m,q)$.
Consequently, a useful approach begins by set a bound on each parameter $L,m,q$, then perform the nonlinear least square estimation for different combinations of parameters. As a result, if there is an occurrence of chaos, there will be a set of positive values of $\lambda$. In this case, the final choice is the largest value.

Once the Lyapunov exponent is estimated, a hypothesis test is performed to assess the confidence level of the prediction. This is an important step since the noise is not completely removed in the estimation of $F$, hence there is some degree of uncertainty in the mapping, especially if the signal to noise ratio is low. The test is based on distributional results for the Lyapunov exponent. Such results rely on the hypothesis of $F$ bounded, its Jacobian is also bounded, the noise variables $\epsilon_t$ are independent and identically distributed, and their probability distribution function is bounded. These conditions yield an asymptotically normal distribution through the central limit theorem.

The above describes the basic mechanics of the estimation of Lyapunov exponents. This method was confirmed with several sets of data, both chaotic and stochastic. For additional details refer to \cite{BenSaida2}.

\section{The Tennessee Eastman plant}

This PWC problem was originally proposed by engineers of the Eastman company as a plantwide control problem challenge for the academic community \cite{DownsVogel}. It is a petrochemical plant given in the form of an algorithm, so mathematical expressions to describe the system are not available. Additional properties are nonlinearity, strong interaction between variables, open-loop instability, and hard operative restrictions over process conditions. The system presents 40 inputs, 50 states, and 41 outputs. The inputs 1 to 12 are related to valves and actuators that it is possible to manipulate to control the plant, while the rest of the inputs represent specific perturbations that any desirable control system should be able to handle. Considering the above features, it is not evident how to choose the CV solely based on process engineering considerations. 

\begin{figure}
  \centering
  \makebox[\textwidth][c]{\includegraphics[width=1.05\textwidth]{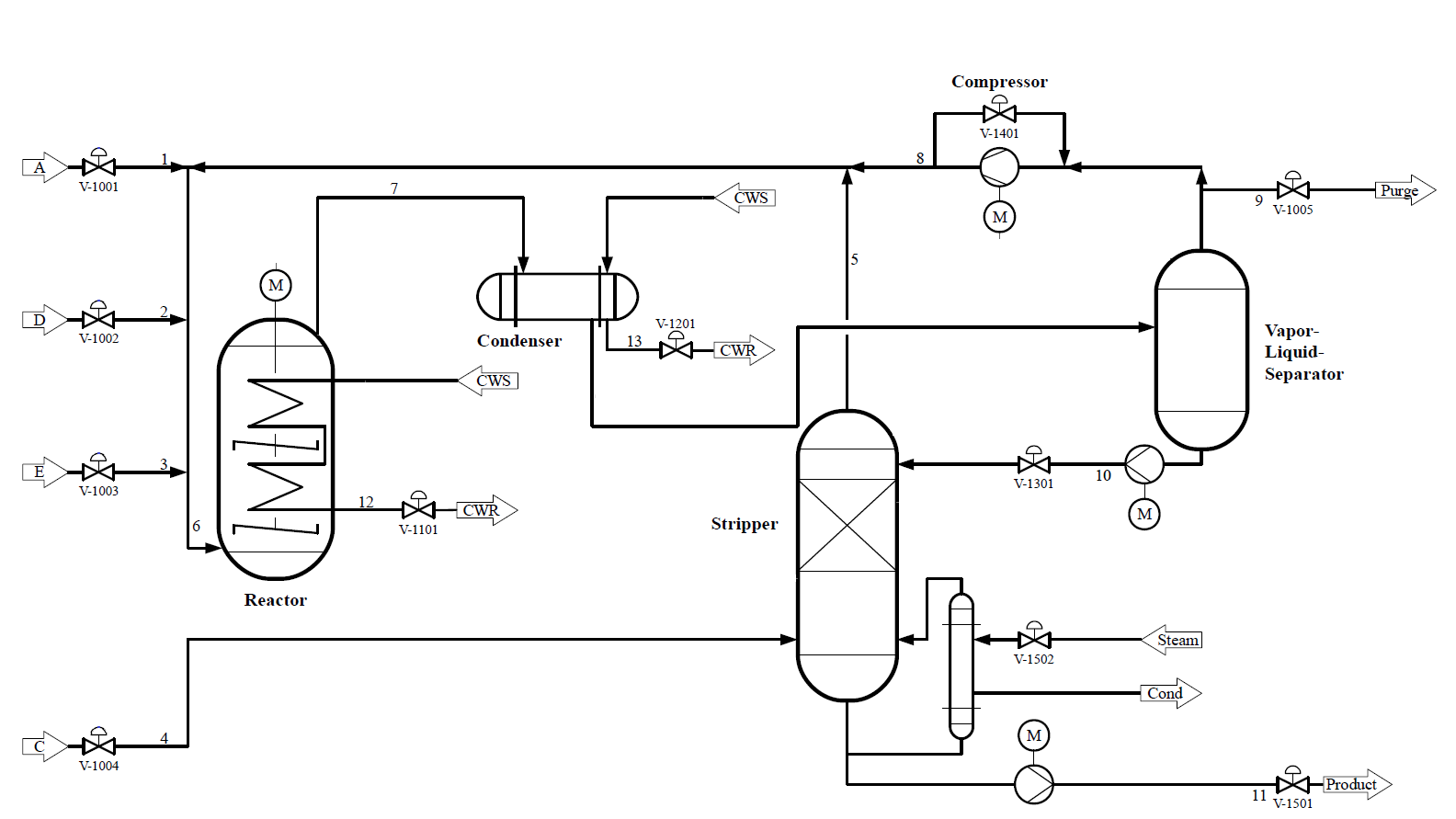}}
  \caption{The Tennessee Eastman Plant.}
  \label{fig:TEplant}
\end{figure}

The Fig. \ref{fig:TEplant} shows a process flow diagram for the plant. The process involves different units: chemical reactor, condenser, separator, compressor, and stripper. Additionally, it has a recycle stream to increase reactant use efficiency. The process is fed with four gas streams, which introduce the reactants A, C, D, E and a small amount of noncondensible component B. Four simultaneous reactions occur at the reactor, all exothermic. They produce two products of interest G and H, and a by-product F. The products are taken from the bottom of the stripper. There is another output stream at the top of the separator which serves as a purge. 

Each processing unit has its own characteristic dynamics. Particularly, the reactor is open-loop unstable, and contains both liquid and vapor phases. The interconnections between the units regarding material streams, momentum transfer, and internal energy, produce a complex behavior for the whole plant. Traditional control system design procedures, based on assembling control subsystems for every unit might fail for complex plants like this one. This is one of the main reasons why plantwide process control is an active research area up to this date.

Even some modern approaches may either fail or underperform against this problem. A distinctive feature of the TE plant is that the linear model is strongly ill-conditioned, meaning that the condition number of the transfer function matrix is extremely high at every frequency of interest. As result, the use of control design methods for selecting CV based on the linear model can be misleading due to numerical issues. An extensive discussion, analysis, and control system design for this plant is found in my Ph.D. thesis \cite{YapurTesisDI}.

\section{Discussion and Results}

Results of the chaos test for the Tennessee Eastman plant output signals are shown in Table \ref{tab:caosSalida}. The first column reveals the identification symbol for the signal, sorted from 1 to 41, while the second column has designations that resemble the type and location of each variable. For instance, TR is the reactor temperature. The third and the fourth columns present the estimated Lyapunov exponent and $p$-value respectively. The first conclusion from this column is that 27 of the output signals present chaotic behavior. However, the degree of certainty varies across the different variables. Notice that some of the Lyapunov exponents are negative. This suggests that in principle, the associated variables would not be chaotic. However, they appear in the table since the $p$-value is relatively small, so there is not enough confidence to reject the null hypothesis $H_0$. On the other hand, some signals have both positive Lyapunov exponents and $p$-values close to unity, confirming a chaotic behavior for those signals. 

For the particular case of the Tennessee Eastman plant, it is worth mentioning that the confidence in the chaoticity may vary across variables with $\lambda>0$ and $p \approx 1$. The reason being a combination of two facts. Firstly, there is a limit in the signal length that is achievable from the open-loop simulation of the plant, since it is programmed to stop whenever any critical variable surpasses any operative limit. Secondly, not all variables have the same sampling time. The signals $y_i$ with $i=23,...,41$ are related to chemical compositions, so they have a higher sampling time compared with the rest of the signals. Consequently, while there are over 600 measures for every output signal $y_i$, with $i=1,...,41$ there are only 11 different numerical values for compositions from $y_{23}$ to $y_{36}$, and 5 different values from $y_{37}$ to $y_{41}$. This leads to less information on the dynamic evolution of the compositions, so conclusions obtained for these signals could be biased by the low volume of actual data acquired. Nonetheless, this fact will be neglected for the remainder of this section, as in real industrial plants there is no shortage of data in general.

\begin{table}[ht]
	\begin{center}
		\caption{Chaos test output}
		\label{tab:caosSalida}
		\begin{tabular}{|cccc||cccc|} 
			\hline 
			Symbol & Abbrev. & $\lambda$ & $p$ & Symbol. &  Abbrev. &  $\lambda$ &  $p$\\ 
			\hline 	
			$y_2$ & F2D    & -4.7226 & 0.1569 & $y_{21}$ & TER   & -4.7226  & 0.1150 \\
			$y_3$ & F3E    & -4.7226 & 0.1569 & $y_{22}$ & TEC   & -4.7226  & 0.0980 \\
			$y_7$ & PR     & -4.7226 & 0.1569 & $y_{24}$ & CBF6R &  0.0081  & 0.6802  \\
			$y_8$ & NR     & -4.7226 & 0.0963 & $y_{26}$ & CDF6R &  0.0014  & 0.5757 \\
			$y_9$ & TR     & -4.7226 & 0.1306 & $y_{28}$ & CFF6R &  0.0014  & 0.5625 \\
			$y_{10}$ & FP    &  0.1029 & 1.0000 & $y_{32}$ & CDP   &  0.2485  & 0.9609 \\
			$y_{11}$ & TS  & -4.7226 & 0.1004 & $y_{34}$ & CFP   &  0.0068  & 0.7721  \\
			$y_{12}$ & NS  & -4.7226 & 0.0558 & $y_{35}$ & CGP   &  0.0150  & 0.9219 \\
			$y_{13}$ & PS  & -4.7226 & 0.1569 & $y_{36}$ & CHP   &  0.0083  & 0.8271 \\
			$y_{15}$ & ND  & -4.7226 & 0.0572 & $y_{37}$ & CDFLD &  0.0037  & 0.6589 \\
			$y_{16}$ & PD  & -4.7226 & 0.1569 & $y_{38}$ & CEFLD &  0.0256  & 0.9990 \\
			$y_{18}$ &TD   & -4.7226 & 0.0826 & $y_{39}$ & CFFLD &  0.3692  & 1.0000 \\
			$y_{19}$ & FVD & -4.7226 & 0.1478 & $y_{40}$ & CGFLD & -4.7226  & 0.0644 \\
			$y_{20}$ & PC  & -4.7226 & 0.1551 &&&&\\
			
			\hline
		\end{tabular}
	\end{center}
\end{table}

All in all, the former results confirm that the plant is a chaotic system, since the variables FP and CFFLD are chaotic almost surely, as shown in Table \ref{tab:caosSalida}. In fact, most of the process output variables exhibit chaotic behavior to some degree, as it is not possible to reject the null hypothesis $H_0$ for 27 of the 41 signals. 

The characterization of chaos enables a simple procedure to prioritize which variables should be controlled, since for real multi-variable systems it is not generally possible, nor desirable, to control every output variable. To perform a reasonable selection, the criterion suggested in this work is to control those variables that are more unpredictable from a chaos theory standpoint. It is expected that this selection will increase the stability and performance of the final control system in a natural manner, since the technique aims to reduce the complexity of the overall system. To this objective, a hierarchical list of potential CV is computed. The list follows from sorting the algorithm output in descending order of the product $p \times \lambda$. This product represents a compromise between the degree of chaotic behavior and the certainty about it. Notice that this sorting automatically sets the variables with $\lambda >0$ above those with $\lambda<0$. However, it is advisable that only variables with $\lambda>0$ should be selected from this list.

It is possible that some processes show no results with $\lambda>0$, yet there are still variables for which the null hypothesis can not be rejected. In this case, an alternative approach is listing the variables in ascending order of $p$-values. This way, the selected CV will be those for which is less probable than $\lambda<0$. What is more, a combination of the two former criteria can be performed by selecting the first variables from the set of $\lambda>0$ with large $p \times \lambda$ values, and then complementing the list with the set of $\lambda \le 0$ with minimum $p$-values.

\begin{table}[ht]
	\begin{center}
		\caption{Controlled variables selection order}
		\label{tab:selecVC}
		\begin{tabular}{|ccccc|} 
			\hline 
			Symbol & Abbrev. & $\lambda$ & $p$ &  $\lambda \times p$\\ 
			\hline 	
			$y_{39}$ & CFFLD & 0.36928 & 1.0000 & 0.36928 \\
			$y_{32}$ & CDP   & 0.24857 & 0.90840 & 0.22580 \\
			$y_{10}$ & FP & 0.10293 & 1.00000 & 0.10293 \\
			$y_{38}$ & CEFLD & 0.02569 & 0.99908 & 0.02566 \\
			$y_{35}$ & CGP & 0.01500 & 0.91616 & 0.01374 \\
			$y_{36}$ & CHP & 0.00833 & 0.82718 & 0.00689 \\
			$y_{24}$ & CBF6R & 0.00815 & 0.69181 & 0.00563 \\
			$y_{34}$ & CFP & 0.00689 & 0.77210 & 0.00532 \\
			$y_{37}$ & CDFLD & 0.00377 & 0.65895 & 0.00248 \\
			$y_{26}$ & CDF6R & 0.00145 & 0.57573 & 0.00084 \\
			$y_{28}$ & CFF6R & 0.00148 & 0.56257 & 0.00083 \\ 
			\hline
		\end{tabular}
	\end{center}
\end{table}

Now, recalling that the Tennessee Eastman plant has up to twelve manipulated variables, one might be interested in selecting a similar amount of output variables to control. To reduce the list, variables with either $p=0$ or $\lambda \le 0$ are filtered out, although this step is not mandatory. According to the above criteria, the set of selected CV is listed in Table \ref{tab:selecVC}. According to these results, CFFLD is the first CV to be selected, followed by CDP, FP, and so forth. Notice that variables with higher $\lambda \times p$ should be associated with more robust control loops. Consequently, the resulting hierarchy of variables should impact on the controller tuning. In extreme cases, robust control methods should be used for these variables.

\section{Conclusions}

The former study shows potential in the characterization of chaotic signals for the selection of controlled variables in a control system design problem. This criterion is especially useful for unstable, highly multivariable plants with complex dynamics. These kinds of features arise naturally in industrial plantwide processes. 

The criteria proposed can be adapted to a wide variety of levels of output-unpredictability. It is also very simple to compute if the recommended algorithm is used. Additionally, it is a data-driven technique, meaning that it relies uniquely on open-loop data from the process, regardless of how complex it is. However, because of the novelty of this work, it is advisable that these criteria should be complemented with additional suitable process considerations, that will be strongly dependent on the process under study.


\bibliographystyle{unsrt}  


\begin{thebibliography}{1}

\bibitem{GonzalesMirandaCaos}
J.M. Gonzales-Miranda.
\newblock Syncronization and control of chaos.
\newblock In {\em Imperial College Press, 2004}.

\bibitem{BenSaida}
A. BenSa\"ida.
\newblock A practical test for noisy chaotic dynamics.
\newblock In {\em SoftwareX, 2015 3-4}, pages 1--5.

\bibitem{Slotine}
J.J.E. Slotine and W. Li.
\newblock Applied Nonlinear Control.
\newblock Prentice Hall, 1991.

\bibitem{Cover2006IT}
	T. M. Cover and J. A. Thomas.
\newblock Elements of Information Theory
\newblock John Wiley, 2006.

\bibitem {Ricker_Archive},
	N. L. Ricker.
	\newblock Tennessee Eastman challenge archive.
	\newblock {\url{http://depts.washington.edu/control/LARRY/TE/download.html}}, 2005
	
\bibitem{Luyben1998}
	W. L.Luyben and B. D. Try'eus and M. L. Luyben.
	\newblock Plantwide Process Control.
	\newblock McGraw-Hill, 1998.
	
\bibitem{Skogestad_PWC_1}
    S. Skogestad.
    \newblock Plantwide Control: Towards a Systematic Procedure
    \newblock European Symposium on Computer Aided Process Engineering, 12, 57-69, 2002.
    
\bibitem{Tewari}
    A. Tewari
    \newblock Modern Control Design.
    \newblock John Wiley and Sons, 2002.

\bibitem{Ellacott}
    S. W. Ellacott and J. C. Mason and I. J. Anderson.
    \newblock Mathematics of Neural Networks.
    \newblock Springer Science, 1997.

\bibitem{BenSaida2}
    A. BenSa\"ida.
    \newblock Noisy chaos in intraday financial data: Evidence from the American index.
    \newblock Applied Mathematics and Computation, 226, 258-265, 2014.

\bibitem{Eckmann}
    J. P. Eckmann and D. Ruelle.
    \newblock Ergodic theory of chaos and strange attractors.
    \newblock The American Physical Society, 57, 617-656, 1985.
    
\bibitem{YapurTesisDI}
    S. Yapur.
    \newblock Diseño de un sistema de control para una planta completa usando controladores fraccionales.
    \newblock UNL, 2021.
    
\bibitem{DownsVogel}
    J. J. Down and E. F. Vogel.
    \newblock A plant-wide industrial process control problem.
    \newblock Computers Chem. Eng., 17, 245-255, 1993.
    

\end{thebibliography}

\end{document}